\documentclass[journal=jacsat,layout=twocolumn,manuscript=article]{achemso}

\usepackage{graphicx}
\usepackage{dcolumn}
\usepackage{bm}
\usepackage[utf8]{inputenc}
\usepackage[T1]{fontenc}
\usepackage{mathptmx}
\usepackage{etoolbox}
\usepackage{float}
\usepackage{chemformula}
\usepackage{graphicx}
\usepackage{natbib}
\usepackage{textgreek}
\usepackage{siunitx}
\usepackage[utf8]{inputenc}
\usepackage{amsmath}
\usepackage{xcolor}
\usepackage[colorlinks=true,linkcolor=blue,citecolor=blue,allcolors=blue]{hyperref}
\usepackage[
]{achemso}
\setkeys{acs}{maxauthors=0}
\setkeys{acs}{etalmode=truncate}
\setkeys{acs}{articletitle = true}

\newcommand{\LMU}{Faculty of Physics, Ludwig-Maximilians-Universität München, Schellingstraße 4, 80799 Munich, Germany}
\newcommand{\WSI}{Walter Schottky Institute and Physics Department, Technical University of Munich, Am  Coulombwall 4a, 85748 Garching, Germany}
\newcommand{\MCQST}{Munich Center for Quantum Science and Technology (MCQST), Schellingstr. 4, 80799 Munich, Germany}
\newcommand{\NIMS}{International Center for Materials Nanoarchitectonics, National Institute for Materials Science, Tsukuba 305-0044, Japan}
\newcommand{\NIMSTwo}{Research Center for Functional Materials, National Institute for Materials Science, Tsukuba 305-0044, Japan}
\newcommand{\KIT}{Physikalisches Institut, Karlsruhe Institute of Technology, Wolfgang-Gaede-Str. 1, 76131 Karlsruhe, Germany}
\newcommand{\Qlibri}{Qlibri GmbH, Maistr. 67, 80337 Munich, Germany}

\author{Florian Sigger}
    \altaffiliation{These authors contributed equally}
    \affiliation{\WSI}
    \alsoaffiliation{\MCQST}
\author{Ines Amersdorffer}
    \altaffiliation{These authors contributed equally}
    \affiliation{\Qlibri}
    \alsoaffiliation{\LMU}
    \alsoaffiliation{\MCQST}
    \email{ines.amersdorffer@physik.lmu.de}
\author{Alexander Hötger}
    \affiliation{\WSI}
    \alsoaffiliation{\MCQST}
\author{Manuel Nutz}
    \affiliation{\Qlibri}
    \alsoaffiliation{\LMU}
    \alsoaffiliation{\MCQST}
\author{Jonas Kiemle}
    \affiliation{\WSI}
    \alsoaffiliation{\MCQST}
\author{Takashi Taniguchi}
    \affiliation{\NIMS}
\author{Kenji Watanabe}
    \affiliation{\NIMSTwo}
\author{Michael Förg}
    \affiliation{\Qlibri}
    \alsoaffiliation{\LMU}
    \alsoaffiliation{\MCQST}
\author{Jonathan Noe}
    \affiliation{\Qlibri}
    \alsoaffiliation{\LMU}
    \alsoaffiliation{\MCQST}
\author{Jonathan J. Finley}
    \affiliation{\WSI}
    \alsoaffiliation{\MCQST} 
\author{Alexander Högele}
    \affiliation{\LMU}
    \alsoaffiliation{\MCQST}  
\author{Alexander W. Holleitner}
    \affiliation{\WSI}
    \alsoaffiliation{\MCQST}
\author{Thomas Hümmer}
    \affiliation{\Qlibri}
    \alsoaffiliation{\LMU}
    \alsoaffiliation{\MCQST}
\author{David Hunger}
    \affiliation{\KIT}
\author{Christoph Kastl}
    \affiliation{\WSI}
    \alsoaffiliation{\MCQST}
    \email{christoph.kastl@wsi.tum.de}

\title[]{Ultra-sensitive extinction measurements of optically active defects in monolayer MoS$_2$}

\keywords{2D materials, helium ion microscopy, defect absorption, cavities, microcavity, absorption microscopy}

\begin{document}

\begin{abstract}
\textbf{We utilize cavity-enhanced extinction spectroscopy to directly quantify the optical absorption of sulfur vacancy defects in MoS$_2$ generated by helium ion bombardment. We achieve hyperspectral imaging of specific defect patterns with a detection limit below 0.01\% extinction, corresponding to a detectable defect density below \SI{1e11}{cm^{-2}}. The corresponding spectra reveal a broad sub-gap absorption, being consistent with theoretical predictions related to sulfur vacancy-bound excitons in MoS$_2$. Our results highlight cavity-enhanced extinction spectroscopy as efficient means for the detection of optical transitions in nanoscale thin films with weak absorption, applicable to a broad range of materials.}
\end{abstract}
The absorption of light in a crystal is a process of fundamental importance in condensed matter physics because it is directly related to the excited state spectrum of the crystal Hamiltonian including many-body interactions, such as excitonic effects in low-dimensional semiconductors \cite{Haug.2009, RefaelyAbramson.2018}. Therefore, direct, sensitive, and quantitative measurements of absorbance are key experiments not only in classical optics, but also for solid-state quantum systems \cite{Lin.2019,Barre.2022}. While absorbance measurements are nowadays in the standard repertoire of experimental characterization techniques for bulk crystals, the absorbance of truly nanoscale systems, such as nanometer-thin films or individual nanoparticles, remains notoriously hard to measure due to their small absorption cross section. A notable exception are atomically thin, two-dimensional van der Waals semiconductors, where reduced dielectric screening and many-body interactions lead to sharp excitonic resonances with up to 15\% peak absorbance in crystals less than a nanometer thick \cite{Li.2014,Wang.2018}. Due to the enhanced light-matter interaction, relatively simple differential absorbance/transmission or spectroscopic ellipsometry measurements can be applied to determine, for example, the fundamental excitation gap or excited state spectrum of these 2D semiconductors \cite{Chernikov.2014,Li.2014,Funke.2016,Wurstbauer.2017}. However, the comparatively high detection limit of these methods (about 0.1\%) still precludes the study of processes with smaller cross sections, in particular excitations with an energy below the excitonic gap. These include, for example, single atomic defects in 2D semiconductors which are promising candidates for deterministically placeable and electronically addressable single photon sources \cite{Chakraborty.2019,MichaelisdeVasconcellos.2022}, as well as long-lived interlayer excitons in van der Waals heterostructures which may realize excitonic many-body states \cite{Jiang.2021,Forg.2019,Sigl.2020}. For atomic defect centers in moderate-gap two-dimensional semiconductors, ab-initio calculations predict a strong mixing between localized defect and extended band states resulting in a quasi-continuum of excitonic states below the free exciton transition \cite{RefaelyAbramson.2018}. Yet, direct experimental probes of the weak defect absorption have remained elusive. Cavity-enhanced extinction microscopy \cite{Mader.2015,Hummer.2016,Gebhardt.2019,Benedikter.2019} can be used to overcome these limitations. This auspicious tool pushes the detection limit of approximately 0.1 \% extinction down to $1 \cdot 10^{-4}$ \% \cite{Mader.2015}. It thereby enables direct probing of absorptive properties of states with weak oscillator strength. Moreover, the absorption can be recorded in a quantitative manner and therefore be directly compared with theoretical predictions.\\
Here, we demonstrate and leverage these advantages to directly measure and image the sub-gap absorption of sulfur vacancy defects in monolayer MoS$_2$. By introducing defect sites in a controlled way using focused helium ion beams \cite{Klein.2021,Mitterreiter.2021}, we are able to locally resolve the impact of different defect densities on the sub-gap absorption with a detection limit down to 0.01\% extinction. Hyperspectral scans reveal a broadband absorption spectrum due to the induced vacancy defects, which is consistent with the theoretically predicted continuum of localized defect excitons \cite{RefaelyAbramson.2018}.
\begin{figure}
\includegraphics{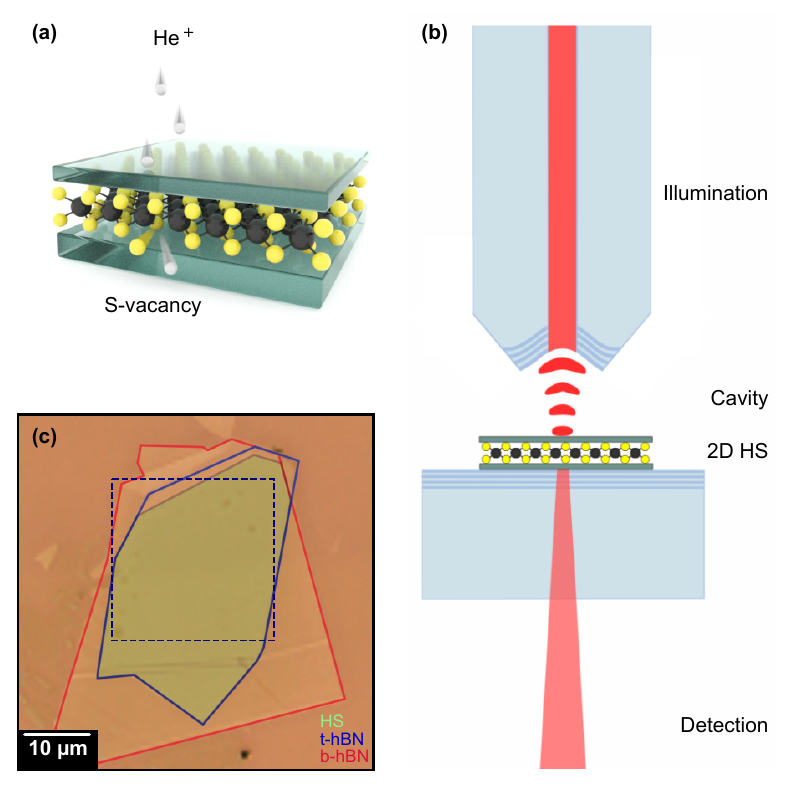}
\caption{\label{Fig1} \textbf{Cavity-based extinction measurements of optically active vacancies in monolayer MoS$_2$.}
(a) Schematic generation of sulfur vacancies in an hBN-encapsulated MoS$_2$ monolayer by He-ion irradiation. (b) A high-finesse cavity is formed between the concave end of a fiber and a plane mirror. Detecting the transmission through the cavity, the extinction of the MoS$_2$ heterostructure (2D HS) is quantified. (c) Optical image of a hBN/MoS$_{2}$/hBN heterostructure on the plane cavity mirror. Green shading indicates the encapsulated region of the MoS$_{2}$. Blue (red) lines denote the top (bottom) hBN layer. The dashed square represents the region of interest for the optical extinction measurements.}
\end{figure}
\hyperref[Fig1]{Figure 1a} depicts schematically the sample geometry and the process of defect generation. Monolayer MoS$_2$ is encapsulated between thin multilayer hBN. Subsequently, defects in the MoS$_2$ monolayer are generated by using a helium ion microscope (HIM)\cite{Klein.2019}. The He-ion beam predominantly induces atomic sulfur vacancies \cite{Mitterreiter.2020,Mitterreiter.2021}. These sulfur vacancies have been shown to act as single photon emission centers with a characteristic low-temperature defect photoluminescence (PL) at approximately $\SI{1.75}{\electronvolt}$ \cite{Klein.2021,Barthelmi.2020}. In order to probe the room temperature optical absorption of the defective MoS$_2$, we integrate the heterostructure into a fiber-based plane-concave scanning cavity microscope (a prototype of Qlibri GmbH) as shown in \hyperref[Fig1]{Figure 1b}. A distributed Bragg reflector (DBR) on a plane quartz substrate is combined with a concave DBR micro-fabricated onto the end facet of a tapered single-mode fiber forming the active volume of a high-finesse microcavity\cite{Mader.2015,Gebhardt.2019,Benedikter.2019}. The mirrors consist of alternating dielectric layers, with SiO$_2$ as the topmost layer. For the present study, the properties of the mirrors were chosen to yield a broadband cavity with an approximate finesse of 20000 (Supporting Information, Figure S1) in the spectral range between \SI{1.55}{\electronvolt} (\SI{800}{\nano\meter}) and \SI{1.77}{\electronvolt} (\SI{700}{\nano\meter}) to match the expected absorption energy of the sulfur defects at room temperature \cite{Klein.2019}. The concave geometry of the micro-mirror laterally confines the cavity mode to a spot size with a radius of approximately $\SI{1.2}{\micro\meter}$ (at half width at half maximum). The cavity is excited via the single-mode fiber using a tunable light source. A specific cavity length is selected by adjusting the distance between the fiber and the mirror with piezo-electric transducers. During measurement, the cavity length is tuned over a fraction of a free spectral range, i.e. the separation between two subsequent cavity modes, and when coming into resonance with the introduced light, the transmission is detected. In a simple picture, this approach enhances very weak absorbance features into a measurable quantity by the multiple round trips of the resonant probe light in the cavity. By scanning the fiber parallel to the plane of the mirror and detecting the transmission through the cavity for a fixed probe energy (wavelength), the extinction, which includes both scattering and absorption of the light in the heterostructure, is mapped out spatially. A complete hyperspectral extinction map is obtained by acquiring such extinction images at different wavelengths.\\
\hyperref[Fig1]{Figure 1c} shows an optical microscope image of the hBN/MoS$_2$/hBN heterostructure integrated into the cavity. The heterostructure was first prepared and then transferred onto the plane cavity mirror by a polycarbonate (PC)-based dry stamping process (for details, see Supporting Information) \cite{Purdie.2018}.\\
\begin{figure}
\includegraphics{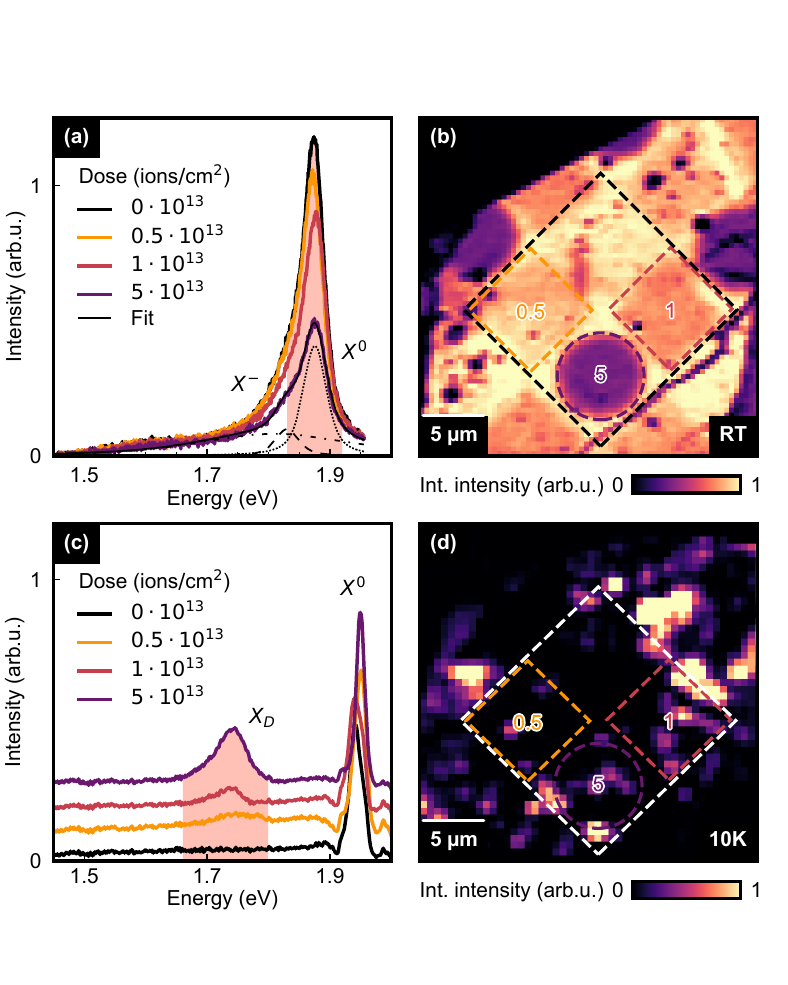}
\caption{\label{Fig2} \textbf{Photoluminescence (PL) of He-ion exposed monolayer MoS$_2$.}
(a) At room temperature, the free exciton emission is reduced with increasing ion dose. The lines denote fits comprising a neutral exciton X$^0$ (dotted), trion X$^-$ (dashed), and a background (dashed-dotted). 
(b) PL image of the irradiated heterostructure integrated from \SI{1.83}{\electronvolt} to \SI{1.92}{\electronvolt} (red shading in a). The exposure pattern (dashed line) consists of a circle (bottom, $5 \cdot 10^{13}$ ions/cm$^2$, labelled 5), a square (right, $1 \cdot 10^{13}$ ions/cm$^2$, labelled 1), a square (left, $0.5 \cdot 10^{13}$ ions/cm$^2$, labelled 0.5), and an unexposed control square (top). The pattern is repeated across the sample. Experimental parameters are $T = \SI{293}{\kelvin}$, cw-excitation, $E_\text{ph}=\SI{2.33}{\electronvolt}$, $P_\text{laser}=\SI{50}{\micro\watt}$, $\text{NA} = 0.9$.
(c) Low-temperature PL spectra averaged across the areas defined in (b). The defect emission at \SI{1.75}{\electronvolt} increases with ion dose.
(d) PL image of defect emission integrated from \SI{1.66}{\electronvolt} - \SI{1.8}{\electronvolt} (red shading in c). Exposed areas show enhanced defect emission. Experimental parameters are $T = \SI{10}{\kelvin}$, cw-excitation, $E_\text{ph}=\SI{2.54}{\electronvolt}$, $P_\text{laser}=\SI{5}{\micro\watt}$, $\text{NA} = 0.42$.}
\end{figure}
In order to systematically study the impact of defect density on the optical absorption, we treated the MoS$_2$ heterostructure with different ion doses ($0.5\cdot10^{13}$ ions/cm$^2$, $1\cdot10^{13}$ ions/cm$^2$ and $5\cdot10^{13}$ ions/cm$^2$). \hyperref[Fig2]{Figure 2a} displays characteristic room temperature PL spectra (cw-excitation, $E_\text{ph}=\SI{2.33}{\electronvolt}$, $P_\text{laser}=\SI{50}{\micro\watt}$, $\text{NA}=0.9$) for each ion dose. The spectrum comprises mainly emission from the neutral exciton X$^0$ (grey dotted line) and the trion X$^-$ (grey dashed line), whose intensities decrease with increasing the ion irradiation dose. The broad background peak (dash-dotted line) can be attributed to a weak PL from the dielectric substrate. The overall decrease of PL intensity is consistent with non-radiative recombination channels due to the ion-induced generation of defect states \cite{Klein.2018}. \hyperref[Fig2]{Figure 2b} shows a confocal PL intensity map (room temperature, spectral range \SI{1.83}{\electronvolt} to \SI{1.92}{\electronvolt}) of the heterostructure after ion irradiation. The exposure pattern (black dashed line) was placed into a homogeneous area with high-quality interfaces (cf. \hyperref[Fig1]{Figure 1c}) as identified by PL and atomic force microscopy (Supporting Information, Figures S2 and S3). The microscale exposure pattern consists of a circular field (bottom, $5 \cdot 10^{13}$ ions/cm$^2$, labelled 5), a square field (right, $1 \cdot 10^{13}$ ions/cm$^2$, labelled 1), another square field (left, $0.5 \cdot 10^{13}$ ions/cm$^2$, labelled 0.5), and an unexposed control field (top). The pattern was repeated across the sample. The exposure patterns are clearly visible in the PL image due to the reduced free exciton emission in agreement with \hyperref[Fig2]{Figure 2a}.\\
\hyperref[Fig2]{Figure 2c} depicts averaged low temperature spectra from each dose field of the irradiation pattern. The PL spectrum (cw-excitation, $E_\text{ph}=\SI{2.54}{\electronvolt}$, $P_\text{laser}=\SI{5}{\micro\watt}$, $\text{NA}=0.42$) additionally exhibits a defect peak X$_\text{D}$ centered at \SI{1.75}{\electronvolt} from unpassivated sulfur vacancies in agreement with previous studies \cite{Klein.2018,Mitterreiter.2021} with a clear correlation between dose and defect emission intensity. Based on the experimentally determined sputtering yield for sulfur vacancies of 0.02 \cite{Mitterreiter.2020}, we can estimate the generated defect density to be in the range of $10^{11}$ vacancies/cm$^2$ to $2 \cdot 10^{12}$ vacancies/cm$^2$ for the used doses. This constitutes a high density regime, where only a broad defect ensemble peak instead of sharp single defect lines can be observed \cite{Klein.2021}. \hyperref[Fig2]{Figure 2d} shows a corresponding confocal PL image integrated over a spectral window (\SI{1.66}{\electronvolt} - \SI{1.8}{\electronvolt}) comprising the central defect emission energy.\\
The fields with the highest doses in the irradiation pattern can be identified by an increased number of bright emission spots. Some of these bright spots correlate with interface inhomogeneities visible in atomic force microscopy. In 2D semiconductors, structural inhomogeneities, e.g. nanoscale bubbles, locally modify the potential landscape, e.g. via strain fields \cite{Darlington.2020,Rosati.2021}. The resulting local potential minima can funnel excitons into regions with enhanced defect emission, explaining the more localized defect emission at low temperatures. Previous studies demonstrated that by carefully tuning the experimental parameters, such as exposure dose and area, single defect, and thereby single photon emitters, can be deterministically created in monolayer MoS$_2$ and even contacted heterostructures to allow electronic switching of the quantum emission \cite{Klein.2021, Barthelmi.2020, Hotger.2021}. Further studies have generalized this approach to electron beam induced emitters as well as to other van der Waals materials, including WSe$_2$ \cite{parto2021defect} and hBN, whereby the latter is promising for room temperature applications \cite{fischer2021controlled,fournier2021position}.
\begin{figure*}
\includegraphics{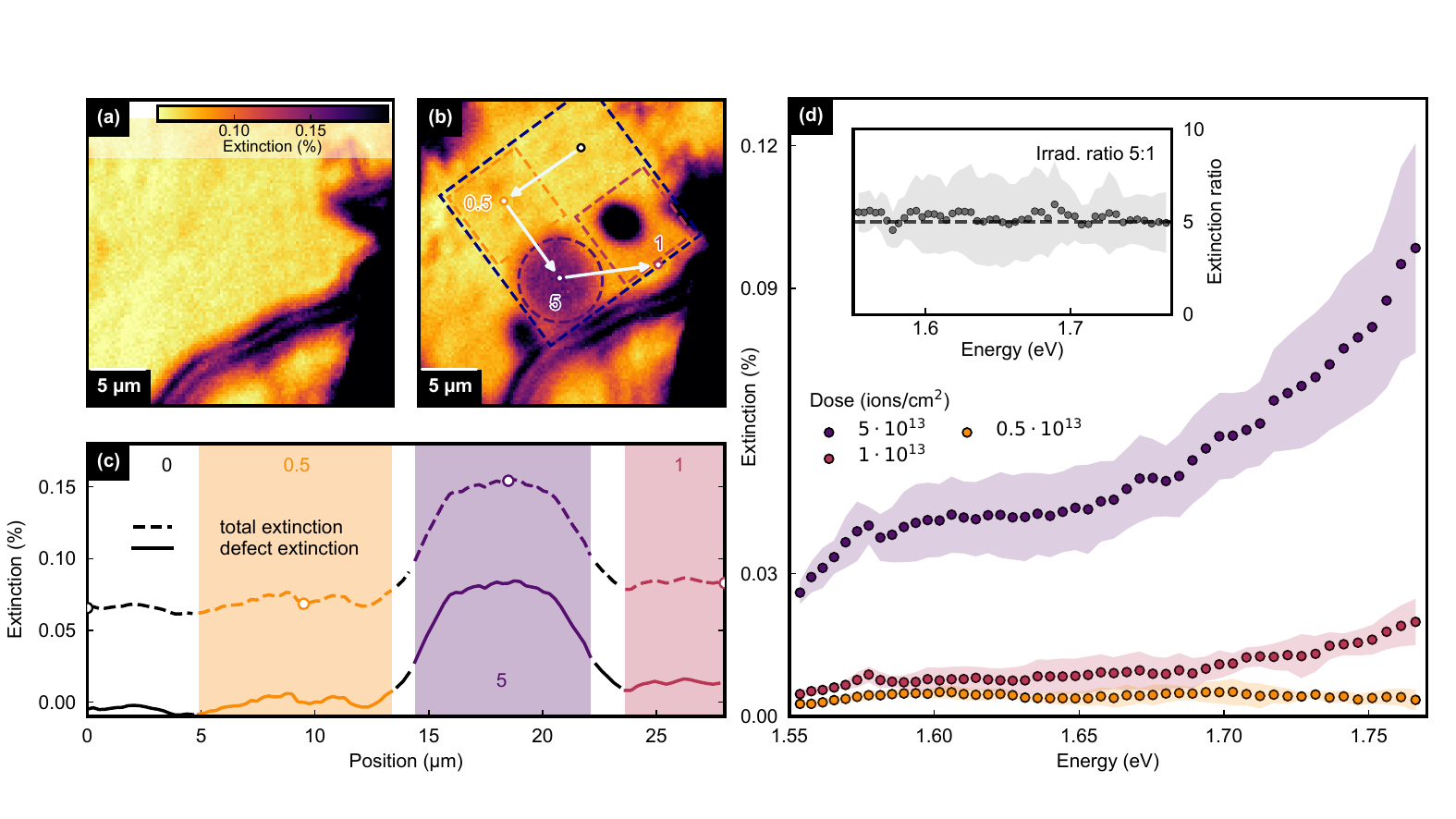}
\caption{\label{Fig3} \textbf{Extinction of ion-induced defects in monolayer MoS$_2$}. Spatial map of extinction (room temperature) measured at $\SI{1.746}{\electronvolt}$ ($\SI{710}{\nano\meter}$) before (a) and after (b) ion irradiation. Before irradiation, the MoS$_2$ monolayer exhibits homogeneous extinction contrast. Structural discontinuities (e.g. edges) give rise to strong scattering and appear dark. After irradiation, the defective areas show a systematically increased sub-gap absorption at the positions of the HIM exposure. (c) Extinction profile cut through all pattern areas for the total extinction (dashed) as displayed in b) and the extinction solely due to the defects (solid) obtained by subtracting the average extinction of the unexposed control field. (d) Spatially averaged extinction spectra corresponding to different defect densities/ion doses. The shaded areas denote the standard deviation of the averaged data points. The inset shows the ratio between the spectra corresponding to a dose of $5 \cdot 10^{13}$ ions/cm$^2$ and $1 \cdot 10^{13}$ ions/cm$^2$. Within the error, the extinction ratio is constant, and it corresponds well to the nominal dose ratio of 5. Due to the absence of structural inhomogeneities in the pattern region, scattering can be neglected and the measured extinction can therefore be interpreted as absorption.}
\end{figure*}
\\
In order to correlate the sub-gap emission to the absorption properties of the induced defects, room temperature extinction scans before (\hyperref[Fig3]{Figure 3a}) and after (\hyperref[Fig3]{Figure 3b}) irradiation were recorded in the cavity-based microscope using illumination with a probe energy of \SI{1.746}{\electronvolt} (\SI{710}{\nano\meter}). Generating the scans benefited from a rather short measurement time of approximately one minute, enabling a rapid sample characterisation.
Sample specific calibration factors, e.g. the effect of the hBN layers on the position of the maximum of the incoming field, are accounted for in the results (Supporting Information, Figure S5).
Before irradiation, the extinction scan shows a spatially homogeneous signal between 0.05\% to 0.1\% across the heterostructure area (\hyperref[Fig3]{Figure 3a}). We find strong extinction values at the edges of the encapsulating hBN and along a crack (Supporting Information, Figure S2) in the MoS$_2$ monolayer, which can be attributed to increased scattering at structural discontinuities of the heterostructure. After irradiation, the exposure pattern becomes visible in the extinction contrast (\hyperref[Fig3]{Figure 3b}).
Note that the strong, localized extinction feature in the field labelled 1 may result from an unintentional structural defect, e.g. laser damage. A similar localized extinction feature is also detected for lower probe energies in the irradiated field labelled 5 (Supporting Information, Figure S4). The affected parts of the irradiated fields are therefore excluded from the following analysis. \hyperref[Fig3]{Figure 3c} displays a line cut across the different fields along the path highlighted in \hyperref[Fig3]{Figure 3b} of the measured extinction (dashed) and the extinction caused solely by the defects (solid). The latter is obtained by subtracting the average extinction in the unexposed reference field. The extinction values increase systematically as the irradiation dose increases, and we can reliably resolve extinction down to 0.01\%. The residual noise stems from the imperfect cavity mirrors and the sample inhomogeneity. In the flat and structurally homogeneous areas of the heterostructure, where scattering is negligible, the signal can be interpreted in terms of a sub-gap absorption due to in-gap defect states of ion-induced sulfur vacancies \cite{RefaelyAbramson.2018,Bretscher.2021b,Mitterreiter.2021}.

\hyperref[Fig3]{Figure 3d} depicts the extinction spectra for the different fields in the exposure pattern. To this end, a series of spatial maps was recorded at different probe energies (Supporting Information, Figure S4), and the average extinction spectrum in each dose field was extracted. Additionally, to ensure that the spectra are not impacted by the use of a specific mode order (determined by the chosen cavity length), the signal was averaged over several cavity lengths in the range between $\SI{5}{\micro\meter}$ to $\SI{7}{\micro\meter}$. The data points denote the corresponding mean values, while the shaded areas denote the standard deviation. All spectra show the extinction caused solely by the defects.

For the exposure fields with doses $1 \cdot 10^{13}$ ions/cm$^2$ and $5 \cdot 10^{13}$ ions/cm$^2$, a plateau of increased extinction can be observed below approximately \SI{1.65}{\electronvolt}. For larger energies, we find a consistent rise in extinction. Importantly, the ratio between the two spectra is constant around a value of 5 (see inset of \hyperref[Fig3]{Figure 3d}), which matches exactly the ratio of the corresponding ion doses. In other words, the change in extinction matches the expected change in ion-induced defect density. Therefore, the spectral shape can be directly interpreted as the absorption spectrum of the defects.
For sulfur vacancies, excitonic many-body calculations predict a continuum of sub-gap states in absorption with the lowest state approximately at around \SI{600}{\milli\electronvolt} below the neutral exciton energy \cite{Mitterreiter.2021}. This is consistent with our observation of an enhanced extinction down to at least \SI{300}{\milli\electronvolt} below the neutral exciton energy, whereby the latter is limited by the stop-band of the used cavity mirrors.
Generally, a direct comparison between defect absorption and defect emission spectra remains challenging. The emission occurs only at low temperatures, since at room temperature interactions with phonons quench the defect PL \cite{Klein.2019}. By contrast, the cavity absorption measurements are currently still limited to room temperature, where the neutral exciton is red-shifted by approximately \SI{50}{\milli\electronvolt} and where a similar red-shift can be expected for the defect transitions\cite{Klein.2018}. Lastly, we quantitatively compare our experimental results to ab-initio predictions for the absorbance of sulfur vacancies in monolayer MoS$_2$. Based on the experimentally determined sputtering yield of 0.02 vacancies per ion from ref.\cite{Mitterreiter.2020}\,, our highest ion dose translates into a defect density of $10^{12}$ cm$^{-2}$ or equivalently a sulfur deficiency of 0.043\%. Calculations at a sulfur deficiency of 2\% yield a defect-induced absorbance of approximately 1\% \cite{Amit.3262022}. By linearly extrapolating the calculated value down to the defect density in our experiments, we arrive at an expected absorption of 0.022\%, which is in the same order of magnitude as the experimental value of 0.042\%.\\
In summary, we have shown the potential of open scanning cavities for the spectroscopy of the vanishingly small sub-gap absorption of 2D nanomaterials that is inaccessible with conventional methods. Specifically, we demonstrated spectrally-resolved cavity-enhanced extinction microscopy on vacancy defects in a MoS$_2$ monolayer with a spatial resolution of approximately $\SI{2.4}{\micro\meter}$. In contrast to modulation based reflectance (or absorbance) measurements \cite{Barre.2022}, cavity-enhanced extinction microscopy yields quantitative and absolute absorption values that may be directly compared to theoretical calculations. Our measured absorption quantitatively confirms theoretical predictions to within a factor of two. The open cavity design of the microscope, where the sample can simply be transferred onto the planar cavity mirror, is ideally suited for the study of 2D materials and layered materials in general. Furthermore, absorption spectroscopy can generally access a wider range of electronic states compared to emission spectroscopy where often only a single dominant decay channel is detected. Specifically for the case of quantum emitters in 2D semiconductors, the optical emission process is observable only at cryogenic temperatures, whereas we demonstrate direct detection of the defect absorbance even at room temperature. Ongoing and future experimental developments include extending the measurement capabilities to cryogenic temperatures to generate a complete picture of the absorption of nanomaterials or defects within them as well as to study the strong-coupling regime in photonic nanosystems.

\begin{acknowledgement}

The work was supported by the Deutsche Forschungsgemeinschaft (DFG, German Research Foundation) within the Priority Programme SPP 2244 2DMP, the Munich Center for Quantum Science and Technology No. (MCQST)-EXC-2111-390814868, e-conversion–EXC No. 2089/1-390776260, the Munich Quantum Valley K6, which is supported by the Bavarian state government with funds from the Hightech Agenda Bayern Plus, the One Munich Strategy Forum – EQAP and the Bundesministerium für Wirtschaft und Klimaschutz (BMWK, Federal Ministry for Economic Affairs and Climate Action) and Europäischer Sozialfonds (ESF, European Social Fund) within an EXIST Transfer of Research (Qlibri, Grant No. 03EFPBY231). A.H. acknowledges support from the European Research Council (ERC) under the Grant Agreement No. 772195, K.W. and T.T. acknowledge support from JSPS KAKENHI (Grant No. 19H05790, 20H00354 and 21H05233).

\end{acknowledgement}

\section*{Conflict of Interest}
The authors have no conflicts to disclose.
\section*{Data Availability Statement}
The data that support the findings of this study are available from the corresponding author upon reasonable request.

\section*{Copyright Information}
This document is the unedited Author's version of a Submitted Work that was subsequently accepted for publication in The Journal of Physical Chemistry Letters, copyright © 2022 American Chemical Society after peer review. To access the final edited and published work see \url{http://pubs.acs.org/articlesonrequest/AOR-UWCKAP2TC3BKZYXK34RT}.

\providecommand{\latin}[1]{#1}
\makeatletter
\providecommand{\doi}
  {\begingroup\let\do\@makeother\dospecials
  \catcode`\{=1 \catcode`\}=2 \doi@aux}
\providecommand{\doi@aux}[1]{\endgroup\texttt{#1}}
\makeatother
\providecommand*\mcitethebibliography{\thebibliography}
\csname @ifundefined\endcsname{endmcitethebibliography}
  {\let\endmcitethebibliography\endthebibliography}{}

\end{document}

\section*{Experimental Methods}

\begin{figure}
\includegraphics{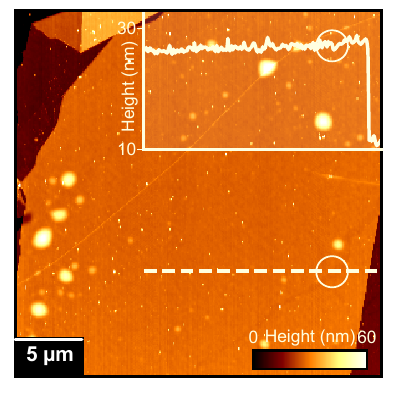}
\caption{\label{SIAFM} Morphology of the heterostructure. Atomic force microscopy image (tapping mode) of the heterostructure discussed in the main manuscript. The sample shows atomically flat and homogeneous topography with the exception of some larger nanobubbles towards the edge of the heterostructures. The irradiation pattern was placed into the central and clean region of the sample. A discontinuity (crack or folding) in the MoS$_2$ monolayer can be identified (see inset), which is consistent with the feature observed in the photoluminescence (\hyperref[SIPLMaps]{Figure S3}) and extinction imaging (\hyperref[SIMaps]{Figure S4}).}
\end{figure}

\begin{figure}
\includegraphics{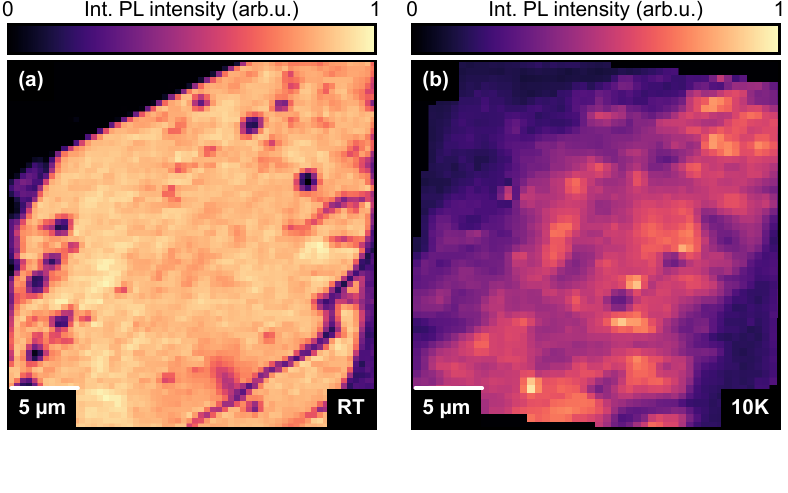}
\caption{\label{SIPLMaps} Integrated PL maps of the neutral exciton X$^0$ emission at (a) room temperature (spectral range from \SI{1.83}{\electronvolt} - \SI{1.92}{\electronvolt}) and (b) low temperature (\SI{10}{\kelvin}, spectral range from \SI{1.92}{\electronvolt} - \SI{1.97}{\electronvolt})}
\end{figure}

\textbf{Heterostructure fabrication.} We prepared fully encapsulated MoS$_2$ heterostructures by a combined approach of different exfoliation/transfer methods. The constituent crystals were first exfoliated using scotch-tape (hBN\cite{Huang.2015}) and viscoelastic dry stamping (MoS$_2$ \cite{CastellanosGomez.2014}) onto SiO$_2$ substrates. Suitable crystals were identified by means of their optical contrast. A hBN thickness of \SI{10}{\nano\meter} to \SI{20}{\nano\meter} was used in this work. After the initial exfoliation, the heterostructure was assembled by a polycarbonate (PC)-based stamping/pick-up process \cite{Purdie.2018}. Finally, the heterostructure was laminated onto the plane cavity mirror at \SI{220}{\celsius} and the residual polymer was dissolved in chloroform. This fabrication method removed possible contaminants between the layers and created smooth and clean interfaces on a scale of several micrometers. Atomic force microscopy (tapping mode) as shown in \hyperref[SIAFM]{Figure S2} confirms large area clean interfaces. Additionally, the homogeneous PL signal for the neutral exciton at room temperature (\hyperref[SIPLMaps]{Figure S3 (a)}) as well as low temperature (\hyperref[SIPLMaps]{Figure. S3} (b)) confirms a high interface quality.

\textbf{Defect generation by He-ion irradiation.} Atomic defects in the encapsulated MoS$_2$ heterostructure were created by focused He-ion irradiation. A Zeiss Orion Nanofab He-ion microscope was used to expose the sample with a beam energy of \SI{30}{\kilo\electronvolt} and a low beam current of $\approx$ \SI{0.1}{\pico\ampere}. The step size between adjacent beam spots was $\SI{5}{\nano\meter}$. The dwell time was adjusted to generate exposure doses of $\SI{0.5e13}{ions/cm^2}$, $\SI{1e13}{ions/cm^2}$, and $\SI{5e13}{ions/cm^2}$.

\textbf{Room temperature PL measurements.} Room temperature photoluminescence measurements were implemented with a $E_\text{ph}=\SI{2.33}{\electronvolt}$ ($\SI{532}{\nano\meter}$) cw-probe laser coupled into a commercial confocal microscope (WITec Alpha 300 R) using a 100x (0.9 NA) objective and a 300 g/mm grating. Low power excitation ($P_{\text{laser}} = \SI{50}{\micro\watt}$) allowed measurements in ambient conditions without apparent beam damage.

\textbf{Low temperature PL measurements.} Low temperature photoluminescence measurements were performed using a He-flow cryostat with a base pressure of approximately \SI{e-6}{\milli\bar} and with bath temperature of approximately \SI{10}{\kelvin}. The cw-probe energy was $E_\text{ph}=\SI{2.54}{\electronvolt}$ ($\SI{488}{\nano\meter}$) at a power of $P_\text{laser}=\SI{5}{\micro\watt}$ through a 50x (0.42 NA) objective. The signal was collected through the same objective and reflection geometry, and were analysed with a dispersive spectrometer (\SI{163}{\milli\meter} focal length, \SI{500}{g/mm} grating) and a cooled CCD.

\begin{figure}
\includegraphics{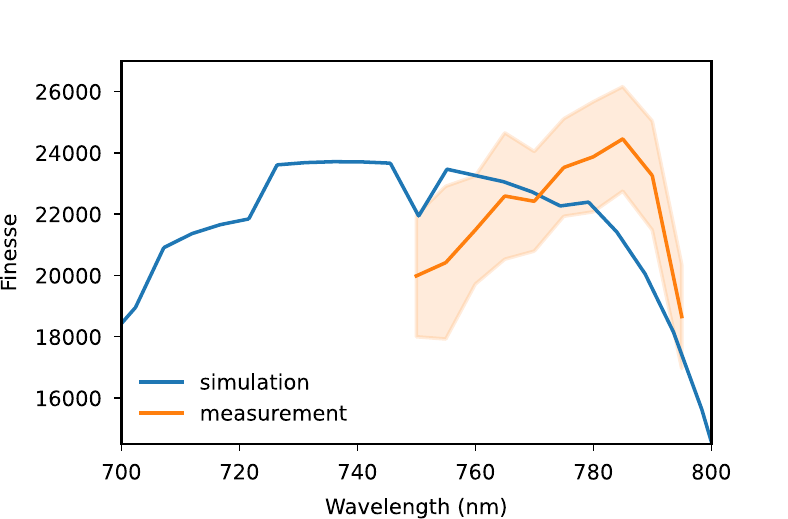}
\caption{\label{SIFinesse} Simulated and measured finesse for wavelengths  $\SI{700}{\nano\meter}$ to $\SI{800}{\nano\meter}$. The shaded area denotes the standard deviation of the averaged datapoints.}
\end{figure}

\begin{figure}
\includegraphics{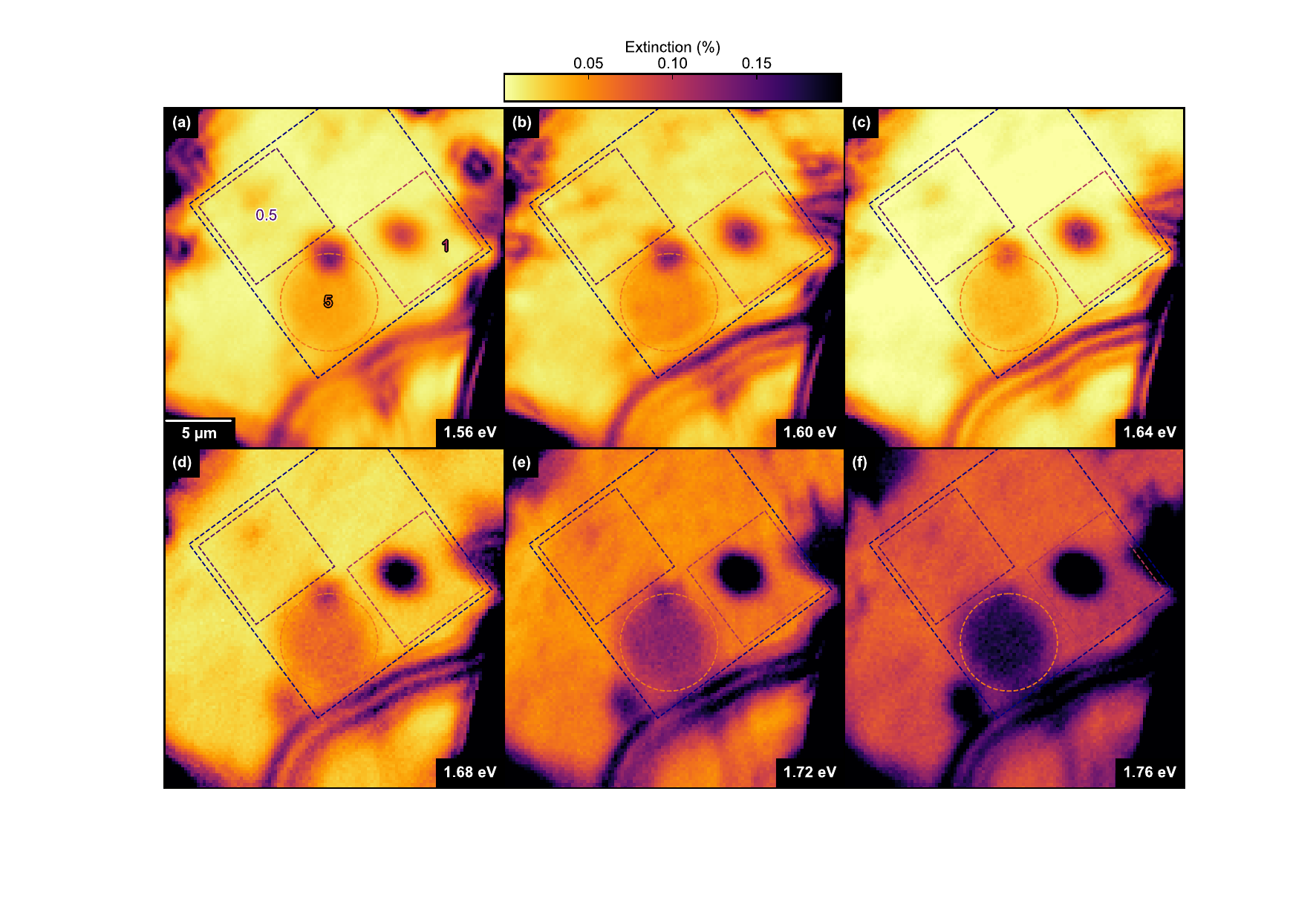}
\caption{\label{SIMaps} Hyperspectral extinction imaging. The images in (a)-(f) display representative extinction images recorded at different probe energies. The extinction spectra presented in the main manuscript were generated from an aligned set of such images.}
\end{figure}

\begin{figure}
\includegraphics{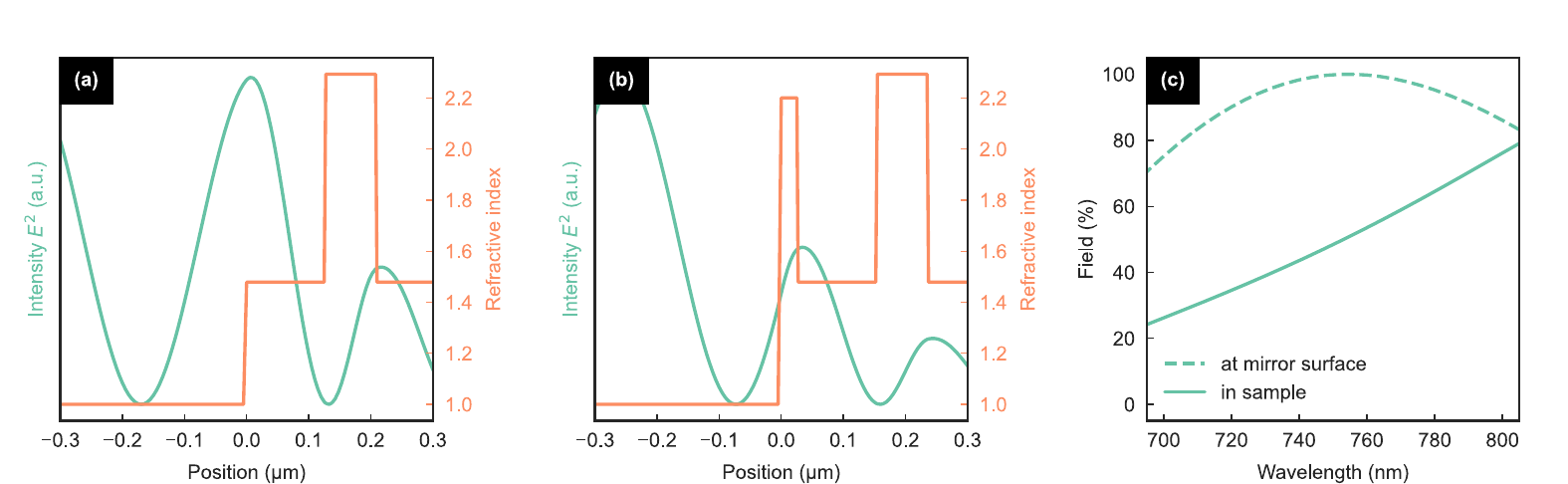}
\caption{\label{SIField} Effect of the sample layer on the field intensity. (a) Refractive indices close to the mirror surface (\SI{0}{\micro\meter}) and the corresponding field intensity. (b) Refractive indices with added hBN layers of the sample. The altered field intensitiy is shown. (c) Comparison of the resulting field intensity at the mirror surface without the sample hBN layers (dashed line) and within the sample layer (solid line) for wavelengths  $\SI{700}{\nano\meter}$ to $\SI{800}{\nano\meter}$.}
\end{figure}

\begin{figure}
\includegraphics{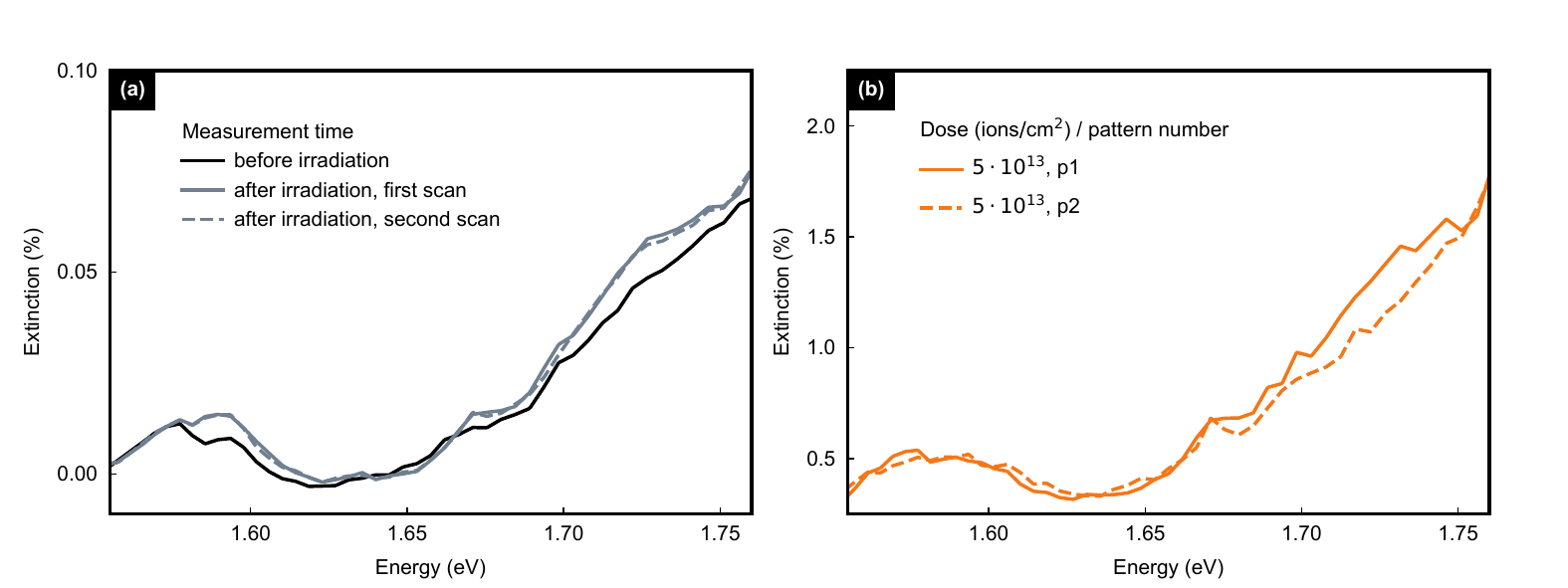}
\caption{\label{SIRep} Repeatability of hyperspectral extinction measurements. (a) Temporal repeatability. Extinction spectra of the heterostructure recorded before the irradiation process (solid black line) and in a non-irradiated area of the heterostructure after the irradiation process, thus after sample handling, (grey line) demonstrate good agreement between scans recorded on different days. Extinction spectra of the heterostructure recorded directly after each other (solid grey and dashed grey lines) show excellent agreement. (b) Spatial repeatability. Extinction spectra recorded on different locations of the sample corresponding to the same ion doses show very good agreement. The spectrum labelled p2 corresponds to the pattern shown in the main manuscript.}
\end{figure}

\textbf{Cavity-enhanced extinction measurements.} All shown cavity-enhanced extinction measurements were carried out at room temperature. The cavity length was kept between $\SI{5}{\micro\meter}$ and $\SI{7}{\micro\meter}$, the probe energy (wavelength) provided by a tunable light source was varied between \SI{1.55}{\electronvolt} and \SI{1.77}{\electronvolt} ($\SI{702}{\nano\meter}$ to $\SI{798}{\nano\meter}$). Simulations of the cavity mirrors and measurements show a finesse of approximately 20000 (\hyperref[SIFinesse]{Figure S1}). The normalised transmission is obtained by forming the ratio of the detected transmission at the sample and at a sample-free region of the planar mirror. The detected values of transmission through a sample-free region of the cavity were in the range of $\SI{0.1}{\micro\watt}$ - $\SI{2}{\micro\watt}$. Taking into account the number of round trips of light in the cavity (finesse $\mathcal{F}$) the extinction of a sample can be inferred from the normalised transmission. Further details, such as the non-linearity of the detectors, the varying background level and the effect of the thickness and refractive index of the sample hBN layers on the field amplitude at the defect region (see \hyperref[SIField]{Figure S5 (a-c)}) were calibrated and incorporated to determine the values of extinction to the best of knowledge. By scanning the fiber relatively to the sample, a two-dimensional map of the extinction can be generated. Varying the probe energy (wavelength) while changing the cavity length to keep the cavity on resonance can thus result in hyperspectral maps of the extinction. Such maps are shown in \hyperref[SIMaps]{Figure S4 (a-f)} with an increasing probe energy for a) to f). Apart from the defect pattern, two localized extinction features in the fields labelled 1 and 5 are visible. The intensity of these extinction features change with probe energy and are assumed to result from unintentional defects of the structure by e.g. laser damage. The extinction caused solely by the defects $E_\text{X}$ (with X corresponding to the different irradiation doses) can be deduced by subtracting the extinction measured for the non-irradiated heterostructure $E_\text{area0}$ from the extinction measured on the irradiated heterostructure $E_\text{areaX}$: $E_\text{X}  = E_\text{areaX} - E_\text{area0}$. In order to investigate the performance of the scanning cavity absorption microscope, analogous measurements were repeated at different sample positions and at different times. \hyperref[SIRep]{Figure S6 (a)} shows the temporal repeatability of the extinction measurements. The extinction spectrum recorded for the heterostructure before ion irradiation is in good agreement with the spectrum recorded after the process of ion irradiation while still on a non-irradiated area. The two grey lines, recorded shortly after each other, show an excellent overlay. \hyperref[SIRep]{Figure S6 (b)} depicts the spatial repeatability of the spectral extinction measurements. The two shown spectra were recorded at two different areas, both irradiated by the dose of $\SI{5e13}{ions/cm^2}$, indicating a good agreement.
\\

\section*{Supplementary data}

\newpage

\providecommand{\latin}[1]{#1}
\makeatletter
\providecommand{\doi}
  {\begingroup\let\do\@makeother\dospecials
  \catcode`\{=1 \catcode`\}=2 \doi@aux}
\providecommand{\doi@aux}[1]{\endgroup\texttt{#1}}
\makeatother
\providecommand*\mcitethebibliography{\thebibliography}
\csname @ifundefined\endcsname{endmcitethebibliography}
  {\let\endmcitethebibliography\endthebibliography}{}